\pdfoutput=1

\documentclass[11pt,twoside,a4paper,cmspaper,final,collab]{cms-tdr}
\def\svnVersion{166416}\def\cmsCernNo{2012-088}\def\cmsCernDate{\today}\def\cmsMessage{Submitted to the European Physical Journal C}

\begin{document}\cmsNoteHeader{FWD-10-014}

\hyphenation{had-ron-i-za-tion}
\hyphenation{cal-or-i-me-ter}
\hyphenation{de-vices}
\RCS$Revision: 152213 $
\RCS$HeadURL: svn+ssh://svn.cern.ch/reps/tdr2/papers/FWD-10-014/trunk/FWD-10-014.tex $
\RCS$Id: FWD-10-014.tex 152213 2012-10-11 11:40:04Z safronov $
\providecommand{\ptmin}{\ensuremath{p_\mathrm{T}^\text{min}}\xspace}
\providecommand{\HERWIGpp} {{\textsc{herwig++}}\xspace}
\ifthenelse{\boolean{cms@external}}{\providecommand{\cmsLeft}{top}}{\providecommand{\cmsLeft}{left}}
\ifthenelse{\boolean{cms@external}}{\providecommand{\cmsRight}{bottom}}{\providecommand{\cmsRight}{right}}
\ifthenelse{\boolean{cms@external}}{\providecommand{\cmsBreak}{\linebreak[4]}}{\providecommand{\cmsBreak}{\relax}}
\cmsNoteHeader{FWD-10-014} 

\title{
Ratios of dijet production cross sections as a function of
the absolute difference in rapidity between jets in proton-proton
collisions at
$\sqrt{s} = 7\TeV$ }
\ifthenelse{\boolean{cms@external}}{\titlerunning{Dijet production as a function of \abs{\Delta(y)}}{}}

\date{\today}

\abstract{
A study of dijet production in proton-proton collisions was performed at
$\sqrt{s}=7\TeV$ for jets with $\pt > 35\GeV$ and $\abs{y}<4.7$ using data collected
with the CMS detector at the LHC in 2010. Events with at least one pair of jets are denoted as ``inclusive''. Events with exactly one pair of jets are called ``exclusive''.
The ratio of the cross section of all pairwise combinations
of jets to the exclusive dijet cross section as a function of the rapidity
difference between jets $\abs{\Delta y}$ is measured for the first time up to
$\abs{\Delta y} =9.2$. The ratio of the cross section for the pair consisting of
the most forward and the most backward jet from the inclusive sample to
the exclusive dijet cross section is also presented. The predictions of the Monte
Carlo event generators \PYTHIA{}6 and \PYTHIA{}8 agree with the measurements.
In both ratios the \HERWIGpp generator exhibits a more pronounced
rise versus $\abs{\Delta y}$ than observed in the data. The BFKL-motivated generators
\textsc{cascade} and \textsc{hej+ariadne} predict for these ratios a significantly stronger
rise than observed.}

\hypersetup{%
pdfauthor={CMS Collaboration},%
pdftitle={Ratios of dijet production cross sections as a function of the absolute difference in rapidity between jets in proton-proton collisions at sqrt(s) = 7 TeV},%
pdfsubject={CMS},%
pdfkeywords={CMS, BFKL, forward jets, dijets, large rapidity intervals}}

\maketitle 

The measurement of inclusive jet production in pp collisions provides an
important testing ground for the Standard Model.
Inclusive jet production is well described at LHC energies, over a wide range in jet transverse momentum and rapidity \cite{CMS-PAP-QCD-10-011,CMS-FWD-10-003}, by calculations at next-to-leading-order (NLO) in perturbative quantum chromodynamics (QCD) using the Dokshitzer-Gribov-Lipatov-Altarelli-Parisi
(DGLAP) approach \cite{Gribov:1972ri, Gribov:1972rt, Lipatov:1974qm, Altarelli:1977zs, Dokshitzer:1977sg} and collinear factorization. The rapidity $y$ is defined as $y = (1/2) \log [(E + p_{z})/(E - p_{z})]$, where $E$ is the jet energy and $p_{z}$
is the component of the jet momentum along the beam axis.
As shown in \cite{Chatrchyan:1343467}, the production of dijets with invariant mass above $165\GeV$ and $\vert y \vert$ less than $2.5$ is also well described by NLO predictions.
When jets are well separated in rapidity, the description of the data becomes worse \cite{CMS-FWD-10-003}.

When the collision energy $\sqrt{s}$ is considerably larger than the hard scattering scale given by the jet transverse momentum, $\pt$, 
the average number of produced jets grows rapidly, along with the phase space available in rapidity. 
This kinematic regime is
expected to be described by the Balitsky-Fadin-Kuraev-Lipatov (BFKL)
evolution \cite{Kuraev:1976ge, Kuraev:1977fs, Balitsky:1978ic} and $k_{\mathrm{T}}$ factorization \cite{Levin:1991ry,Catani:1994sq,Collins:1991el}.
An effective theory has been developed which
describes strong interactions in this kinematic domain \cite{lipatov97}.

The ratio of the dijet production cross section in ``inclusive'' events to that in ``exclusive'' events, \cmsBreak
$R^{\text{incl}}\allowbreak=\allowbreak\sigma^{\text{incl}}/\sigma^{\text{excl}}$, 
as a function of the rapidity separation $\abs{\Delta y}$ between two jets, is a sensitive probe of effects beyond collinear factorization \cite{Kim:1995zu}.
Only jets with transverse momenta
above a minimal value of $\ptmin$ are considered. Events with at least
one pair of jets  are denoted as ``inclusive". Events with exactly one
pair of jets are called ``exclusive".
In the inclusive case, the rapidity separation is evaluated for each pairwise combination of jets above threshold \cite{Kim:1995zu}.
Mueller-Navelet jet pairs \cite{Mueller:1986ey} are a subset of the inclusive dijet class. In
this case only the jet at highest rapidity (i.e. most forward) and that
at lowest rapidity (most backward) are considered. 
At low $\abs{\Delta y}$, the inclusive dijet ratio, $R^{\text{incl}}$, is always larger than the corresponding Mueller-Navelet dijet ratio, $R^{\mathrm{MN}} = \sigma^{\mathrm{MN}}/\sigma^{\text{excl}}$, because all the jets in the rapidity interval between the Mueller-Navelet jets contribute to the inclusive cross section.

From the theoretical point of view, an advantage of the ratios $R^{\text{incl}}$ and  $R^{\mathrm{MN}}$ with respect to the individual dijet production cross sections is that the influence of the uncertainty of the parton distribution functions is greatly reduced; in addition, the ratios are particularly sensitive to the parton radiation pattern \cite{Kim:1995zu}.
At large enough energies, the parton subprocesses involve a large number
of partons with comparable transverse energies.
Such subprocesses, governed by BFKL evolution, lead to an increase of the ratios
with increasing $\abs{\Delta y}$ \cite{Kim:1995zu, Mueller:1986ey, Andersson:2002cf}.
Experimentally, an additional advantage of the $R^{\text{incl}}$ and $R^{\mathrm{MN}}$ ratios defined above is the cancellation of most of the systematic uncertainties affecting both numerator and denominator.

Earlier measurements of Mueller-Navelet jets with $\abs{\Delta y}$ up to 5 were
made at the Tevatron by the D0 experiment \cite{Abachi:1996et,Abbott:1999ai}. D0 did not find
indications of BFKL effects in the azimuthal decorrelation data \cite{Abachi:1996et};
however, a stronger than expected $\sqrt{s}$ dependence of the dijet
production cross section was observed \cite{Abbott:1999ai} when comparing the data at
$\sqrt{s}=630$ and $1800\GeV$ for dijets with large rapidity separation. 
The ATLAS Collaboration recently studied various dijet production cross section ratios at $\sqrt{s}=7\TeV$ for $\abs{\Delta y}< 6$ \cite{ATLASrat}; the results are roughly in agreement with the DGLAP predictions.
At the HERA $ep$ collider, deviations from DGLAP were observed in studies of forward jets by the ZEUS \cite{zeusFw0, zeusFw} and H1 \cite{h1Fw0, h1Fw} Collaborations. However, no compelling evidence for BFKL effects was found.

The component of the CMS detector \cite{cmsexp} most relevant for this analysis is the calorimeter system extending to pseudorapidities $\abs{\eta} = 5.2$, where $\eta=-\log[\tan(\theta/2)]$, and $\theta$ is the polar angle relative to the anticlockwise proton beam direction. The crystal electromagnetic calorimeter (ECAL) and the brass/scintillator hadronic calorimeter (HCAL) extend to pseudorapidities $\abs{\eta} = 3.0$. The HCAL cells map to an array of ECAL crystals to form calorimeter towers projecting radially outwards from 
the nominal interaction point. The pseudorapidity region $3.0<\abs{\eta}<5.2$ is covered by the hadronic forward (HF) calorimeter, which consists of steel absorber wedges with embedded radiation-hard quartz fibers, oriented parallel to the beam direction. The calorimeter towers in the barrel region have segmentation of $\Delta\eta \times \Delta \phi = 0.087 \times 0.087$, becoming progressively larger in the endcap and forward regions ($\Delta\eta \times \Delta \phi = 0.175 \times 0.175$ at $\eta \sim 4.5$).

The CMS trigger system consists of a hardware Level-1 trigger and a software high-level trigger. Jets formed online by the trigger system use ECAL, HCAL and HF inputs for energy clustering and are not corrected for the jet energy response. The minimum-bias trigger was defined as the coincidence of a signal from one of the two beam scintillation counters covering the range $3.23<|\eta| <4.65$ and a signal from one of the beam pick-up timing devices.

The data were collected in 2010, when LHC collided protons at $\sqrt{s}=7\TeV$. Various triggers were used to cover the largest possible range in
rapidity separation between jets. Dijets with moderate  $\abs{\Delta y}$ were
selected by a single-jet trigger with a threshold on the raw
(uncorrected) jet transverse momentum of $15\GeV$. This trigger was
significantly prescaled as the instantaneous luminosity increased, and the effective integrated luminosity recorded with it is $\simeq33\nbinv$.

Dijet events with large rapidity separation are rare. Therefore, a
dedicated trigger for forward-backward dijets was developed. This forward-backward-dijet trigger
selects events with two jets in opposite hemispheres and $\abs{\eta}>3.0$, and
jet raw transverse momentum $\pt > 15\GeV$. It was operated with moderate
prescaling, and the effective integrated luminosity recorded with it is
$\simeq5\pbinv$. This allowed to collect a number of dijet events at high $\abs{\Delta y}$
values more than 100 times larger than with the single-jet trigger.

The trigger efficiency was measured by means of a control sample selected
with the minimum-bias trigger. The single-jet
trigger was found to be 100\%
efficient for dijets with corrected $\pt > 35\GeV$. The single-jet
trigger was also used for the determination of
the efficiency of the
forward-backward-dijet trigger. The latter was 100\% efficient
for dijets with $\pt > 35\GeV$.

Jets were reconstructed offline from the energy depositions in the calorimeter towers,
clustered with the anti-k$_{T}$ algorithm \cite{AntiKT, AntiKTimpl} with a distance parameter $R=0.5$. In the reconstruction process, the contribution from each tower was assigned a momentum, the absolute value and the direction of which were given by the energy measured in the tower, and the coordinates of the tower, respectively. The raw jet energy was obtained from the sum of the tower energies, and the raw jet momentum from the vectorial sum of the tower momenta. The raw jet energies were then corrected to establish a uniform relative response of the calorimeter in $\eta$ and a calibrated absolute response in transverse momentum \pt \cite{JME-10-011pub}. The jet energy resolution for calorimeter jets with $\pt \sim 35\GeV$ is about $22\%$ for $\vert \eta \vert <0.5$ and about $10\%$ for $4< \vert \eta \vert <4.5$ \cite{JME-10-003}. The uncertainty of the jet energy calibration for jets with $\pt \sim 35\GeV$ depends on $\eta$ and is $\simeq 7-8\%$ \cite{JME-10-011pub}. 

In the offline analysis, at least one well-reconstructed primary vertex is required to be present within $\pm24$~cm of the nominal interaction point along the beamline \cite{TRK-10-005}.
In order to reduce the sensitivity to
overlapping pp collisions (the so-called ``pile-up" events), events with
only one primary vertex reconstructed within the luminous region were used
for the measurement.

Loose jet quality cuts \cite{JME-09-008} were used to suppress the
effect of calorimeter noise.
Events with at least two jets with $\pt > 35\GeV$ and $\abs{y}< 4.7$ were
selected; only jets satisfying these criteria were used for the analysis.
All pairwise combinations of jets from the selected events entered the inclusive
distribution. For studies of Mueller-Navelet jets, only the pair
consisting of the most forward and the most backward jet was considered. The
exclusive dijet sample is a subset of the inclusive and Mueller-Navelet
samples, and consists of events where exactly one pair of jets is
found. The measured observables, $R^{\text{incl}}$ and $R^{\mathrm{MN}}$, are defined as the ratios
of the yield of inclusive or Mueller-Navelet dijets to the
yield of exclusive dijets in a specific $\abs{\Delta y}$ bin, respectively.

Detector effects were accounted for by applying bin-by-bin corrections
derived from Monte Carlo (MC) simulations. Simulated events produced with
the generators \PYTHIA{}6 (version 6.422) tune Z2 \cite{Pyt, PytZ1, PytZ2} and \HERWIGpp \cite{Hw++}
(version 2.4.2 with default settings) were passed through the full CMS detector
simulation based on the \GEANTfour package~\cite{geant4}, and were input to the same
event-reconstruction program as used for the data. To quantify
detector effects, the distributions obtained from detector-level
quantities were compared to the distributions obtained at the level of stable particles (lifetime $\tau$ such that $c\tau > 10\unit{mm}$).
The ratios of the stable-particle level and detector-level quantities in a
given bin were used to correct the data.

Because of the finite \pt and $y$ resolutions, the detector-level distributions deviate from the corresponding ones at stable-particle level. Events
can migrate to and from the exclusive or the inclusive samples because of
fluctuations of the measured transverse jet momentum around the \ptmin threshold. The amount of these migrations was estimated not to exceed
20\%. Similarly, the $\abs{\Delta y}$ value measured at detector level may fall into a different $\abs{\Delta y}$ bin compared to the stable-particle level; this effect, for the present data, is typically around 5-10\% and reaches 15-25\% at most.
The influence of these
migrations on the measured ratios $R^{\text{incl}}$ and $R^{\mathrm{MN}}$ is minimal as the effects for numerator and denominator are similar.
The uncorrected $R^{\text{incl}}$ and $R^{\mathrm{MN}}$ ratios are reasonably well reproduced by the \PYTHIA{}6 events and less well by \textsc{herwig}++ events, both of which have been passed through the detector simulation.
The $\pt$ and $\eta$ distributions of the jets at the detector level for both MC generators agree with the data within the jet energy scale uncertainty (see also \cite{CMS-FWD-10-003}). 
The correction factors were therefore obtained with \PYTHIA{}6, while their model
dependence was estimated from the difference between
the \PYTHIA{}6 and \HERWIGpp corrections.

The following sources of systematic effects were considered:
\begin{enumerate}
\item Uncertainty of the jet energy calibration. The uncertainty of the measurement was estimated by shifting the jet energy scale (JES) by the \pt- and $\eta$-dependent uncertainties derived in \cite{JME-10-011pub}.
The resulting variation of the measurements does not exceed 4.2\% for $R^{\text{incl}}$ and 3.8\% for $R^{\mathrm{MN}}$.
\item  Uncertainty of the corrections for detector effects.
\begin{enumerate}
\item Uncertainty due to model dependence of the correction factors. As
discussed above, correction factors were determined by using \HERWIGpp and \PYTHIA{}6, and corrected measurements were obtained for each case. The difference in the results was taken as a measure of the model dependence of the correction factors; it does not exceed 3.4\% for $R^{\text{incl}}$ and 3.3\% for $R^{\mathrm{MN}}$.
\item Uncertainty related to the quality of the MC description of the jet \pt and $y$ resolutions.
These resolutions were modified by $\pm10\%$ in the simulation as recommended in \cite{JME-10-003}.
A conservative estimate of the uncertainty associated to this effect does not exceed $\pm1.0\%$.
\end{enumerate}
\item An additional systematic effect may originate from the extra
energy and jets due to pile-up collisions. As noted earlier, the
measurement was restricted to events with only one reconstructed
primary vertex to reduce the impact of pile-up. However, a contribution from interactions undetected because of vertex reconstruction inefficiency
may still be present.
By comparing data taken at different instantaneous luminosities, an upper limit of 1.3\% was estimated for this effect.
The value was obtained for pairs of jets in the forward region, where the impact of
pile-up on jet reconstruction is more significant~\cite{JME-10-011pub}; the effect is
thus expected to be smaller for jets at central rapidities.

\end{enumerate}

The total systematic uncertainty was calculated as the quadratic sum of the individual uncertainties.
 The total uncertainty is $\abs{\Delta y}$ dependent but always smaller than 5.6\% for $R^{\text{incl}}$ and 4.8\% for $R^{\mathrm{MN}}$ (Table 1).

\renewcommand{\arraystretch}{2}

\begin{table*}[htbH]
\begin{center}
\caption{ Sources of systematic effects and associated uncertainties. The ranges correspond to the variation of the uncertainty with $\abs{\Delta y}$. For different uncertainty sources, the minimum and maximum values may correspond to different $\abs{\Delta y}$ bins. For asymmetric uncertainties the upper and lower limits are shown. 
}\label{tabunc}
\begin{tabular}{|c|c|c|}
\hline
source & $R^\text{incl}$ uncertainty $(\%)$ & $R^{\mathrm{MN}}$ uncertainty $(\%)$ \\
\hline
jet energy scale & {\Large $^{+ (2.2-4.2)}_{-(1.0-3.0)}$} & {\Large $^{+(0.2-3.8)}_{-(0.2-2.3)}$} \\
uncertainty of detector corrections &  $\pm(1.5-3.5)$ & $\pm(0.1-3.4)$ \\
pile-up & $<1.3$ & $<1.3$ \\
\hline
total & {\Large $^{+ (2.8-5.6)}_{-(2.7-4.5)}$} & {\Large $^{+(0.2-4.8)}_{-(0.2-3.7)}$} \\
\hline
\end{tabular}
\end{center}
\end{table*}

The measured ratios, corrected for detector effects, were compared to the predictions of several MC generators at the stable-particle level. A DGLAP leading-order parton shower approach is used in \PYTHIA{}6 (version 6.422) tune Z2 \cite{PytZ2}, \PYTHIA{}8 (version 8.145) \cite{Pyt8} tune 4C \cite{Corke:2010yf} and \HERWIGpp (version 2.5.1) tune UE-7000-EE-3 \cite{Hw++,Hw++25}. 
For the simulation of the non-perturbative fragmentation, \PYTHIA{} uses the Lund string model, and \HERWIGpp
the cluster fragmentation approach. 
The tunes mentioned above include multiple parton interactions (MPI) as a part of the underlying event (UE) modeling.
The dijet observables might be affected by MPI through jets that do not originate from the same hard interaction.
The effect was estimated by switching off the MPI options in \PYTHIA{}6 and \HERWIGpp; no significant change in the predictions was observed.

The Monte Carlo generators \textsc{cascade} (version 2.2.03) \cite{CASCADE2} and \textsc{hej} (version 1.3.2) \cite{HEJ} are motivated by the leading-logarithmic BFKL approach and incorporate parts of a next-to-leading logarithmic approximation. 
The \textsc{hej} generator produces parton-level jets; the
corresponding showers were produced with the \textsc{ariadne} program \cite{ariadne}. The \textsc{hej+ariadne} package \cite{hejAriadne} version 0.99b, consisting of \textsc{hej} 1.3.2 and \textsc{ariadne} 4.12, was used.

\begin{figure}[hbtp]
  \begin{center}
     \includegraphics[width=0.49\textwidth]{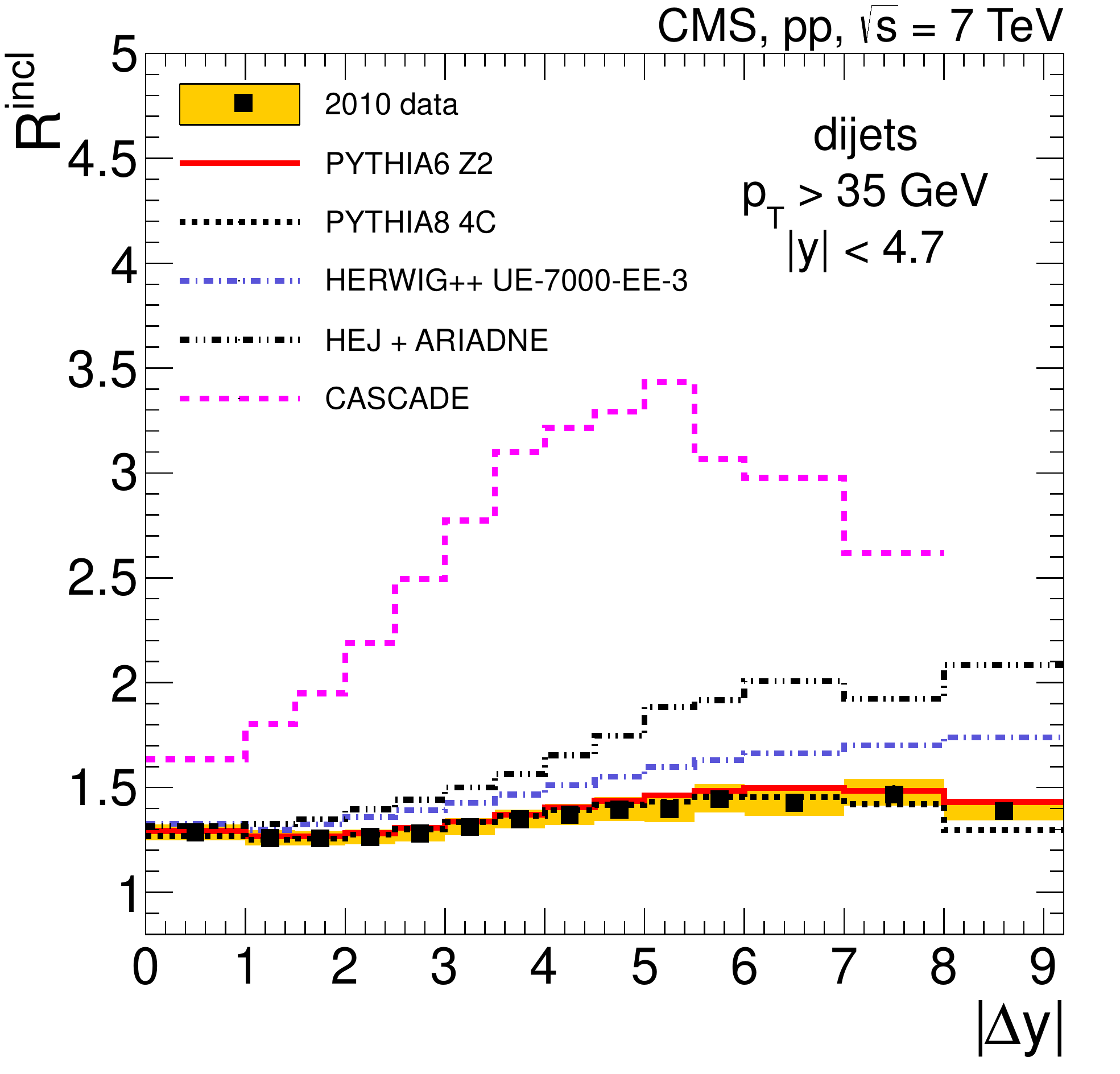}
     \includegraphics[width=0.49\textwidth]{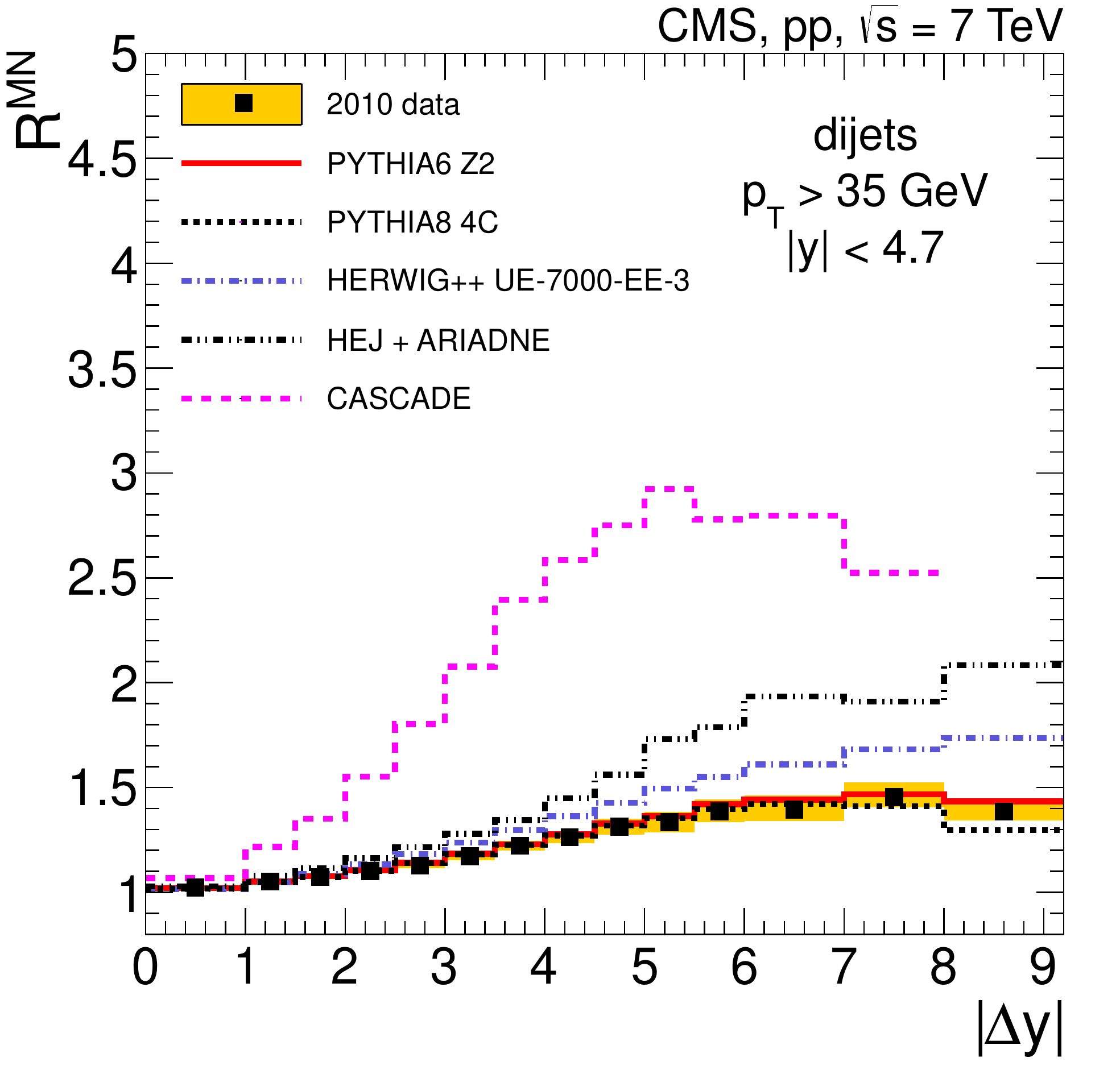}
    \caption{
Ratios of the inclusive to exclusive dijet cross sections as a function of the rapidity
separation $\abs{\Delta y}$ between the two jets, $R^{\text{incl}}$ (\cmsLeft\ panel) and $R^{\mathrm{MN}}$ (\cmsRight\ panel), compared to the predictions of the DGLAP-based MC generators
\PYTHIA{}6, \PYTHIA{}8 and \HERWIGpp, as well as of \textsc{cascade} and \textsc{hej+ariadne}, which incorporate elements of the BFKL approach.
The shaded band indicates the size of the total systematic uncertainty of the data.
Statistical uncertainties are smaller than the symbol sizes.
Because of limitations in the \textsc{cascade} generator, it was not possible to obtain
a reliable prediction for $\abs{\Delta y} > 8$.
}
    \label{fig:ex5}
  \end{center}
\end{figure}
\begin{figure}[hbtp]
  \begin{center}
     \includegraphics[width=0.49\textwidth]{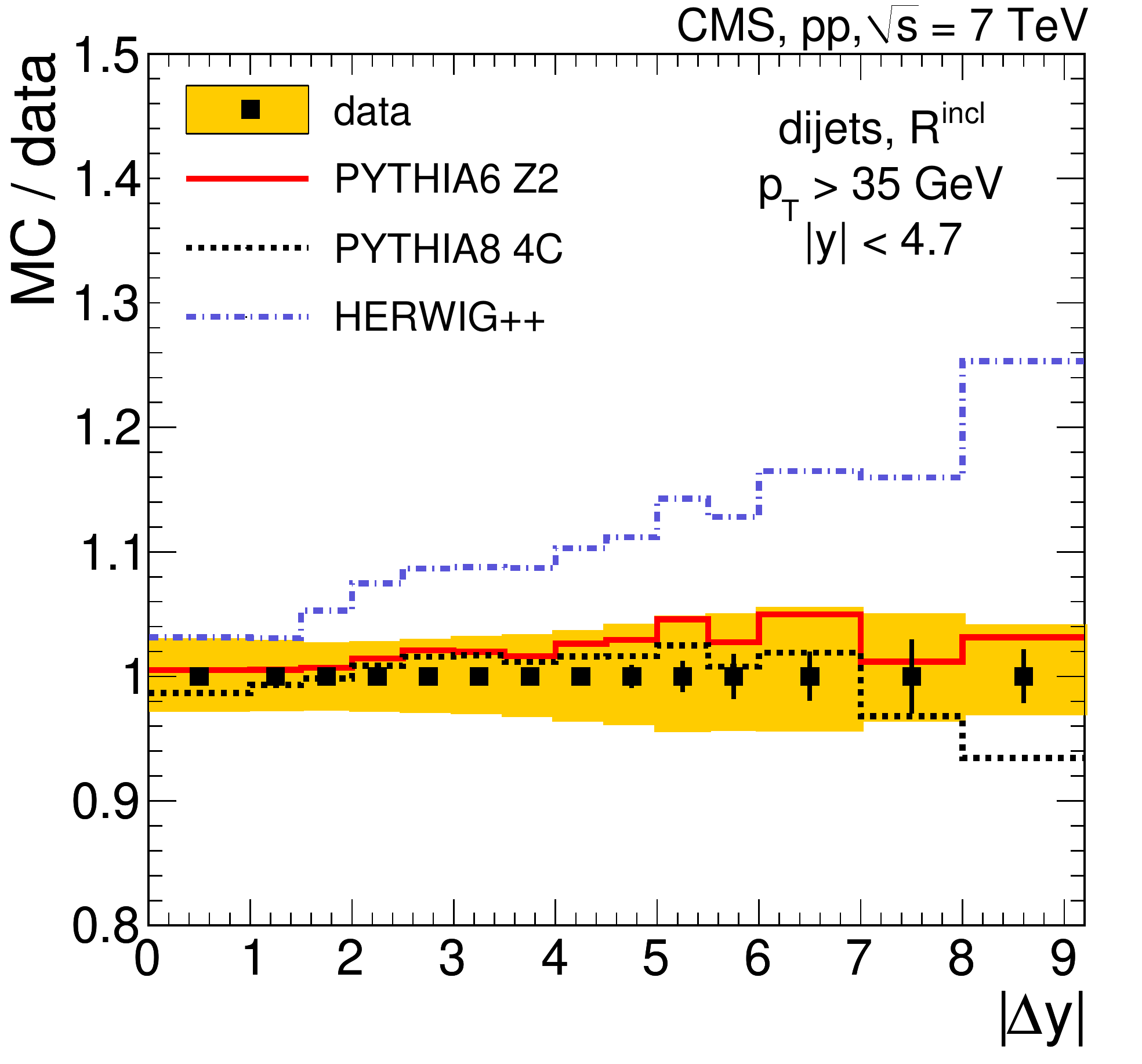} 
     \includegraphics[width=0.49\textwidth]{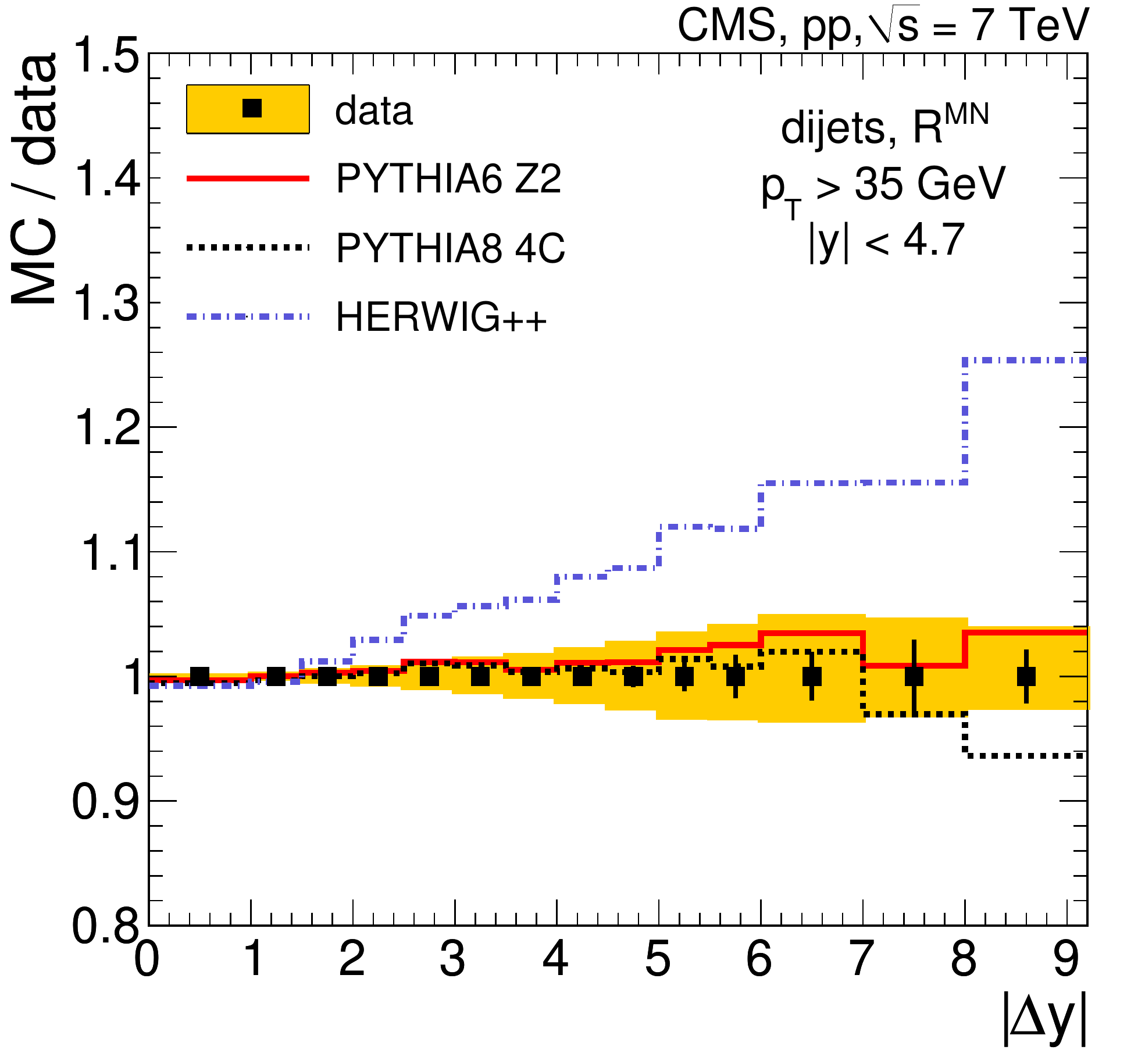} 

    \caption{
Predictions for $R^{\text{incl}}$ (\cmsLeft) and $R^{\mathrm{MN}}$ (\cmsRight) from DGLAP-based
MC generators presented as ratio to data corrected for detector
effects. Both BFKL-motivated generators \textsc{cascade} and \textsc{hej+ariadne} (not shown)
lead to a MC/data ratio well above unity.
The shaded band indicates the size of the total systematic uncertainty of the data while statistical uncertainties are shown as bars.
}
    \label{fig:ex6}
  \end{center}
\end{figure}
The ratio $R^{\text{incl}}$ of inclusive to exclusive dijet production as a function of $\abs{\Delta y}$ is presented in Fig.~\ref{fig:ex5} (\cmsLeft\ panel). On average the inclusive cross section
is 1.2 -- 1.5 times larger than the exclusive cross section.
The ratio $R^{\text{incl}}$ grows with increasing $\abs{\Delta y}$, as expected because of the larger phase space for hard parton radiation.
At the highest $\abs{\Delta y}$, $R^{\text{incl}}$ is expected to decrease because energy-momentum conservation suppresses the emission of extra jets.
The $\abs{\Delta y}$ value where $R^{\text{incl}}$ starts to decrease varies from one MC generator to another, as can be seen in Fig.~\ref{fig:ex5}.

The predictions  from \PYTHIA{}6  and \PYTHIA{}8 agree with the measurement.
\HERWIGpp overestimates the ratio $R^{\text{incl}}$ at medium and large rapidity intervals.
A detailed comparison between the data and the predictions of the DGLAP-based MC generators is presented as a ratio in Fig.~\ref{fig:ex6} (\cmsLeft\ panel).
It was checked explicitly that the results obtained from \PYTHIA{}6 and \HERWIGpp at parton level are close to the corresponding ones at stable-particle level. The different behaviour of \PYTHIA{}6 and \HERWIGpp is also observed at parton level.

The ratio $R^{\text{MN}}$
and the corresponding MC to data ratio are presented in
the \cmsRight\ panels of Figs. ~\ref{fig:ex5}--\ref{fig:ex6}.
At large $\abs{\Delta y}$, $R^{\mathrm{MN}}$ approaches $R^{\text{incl}}$ as extra jet radiation contributing to $R^{\text{incl}}$ tends to concentrate at moderate rapidities.
The quality of the predictions of the DGLAP-based MC generators for $R^{\mathrm{MN}}$ is similar to those for $R^{\text{incl}}$.
The MC generators \textsc{cascade} and \textsc{hej+ariadne} 
considerably overestimate the measurements of both $R^{\text{incl}}$ and $R^{\mathrm{MN}}$. 

The ATLAS Collaboration measured several observables for dijets as functions of the rapidity separation between jets  \cite{ATLASrat}.
One is the fraction of events without jets with $\pt$ above $\pt^{\mathrm{veto}}$ in the rapidity interval between the most forward and the most backward jets, which is equal to $1/R^{\mathrm{MN}}$. 
A difference between data and the parton-level HEJ prediction at large $\abs{\Delta y}$ for $70 < \left< \pt \right> < 90\GeV$, $\pt^{\mathrm{veto}}=\left< \pt \right>$ and $\abs{\Delta y}< 6$ was reported (Fig. 8 from \cite{ATLASrat}), which is in qualitative agreement with the result presented here.

To conclude, the first measurement of the ratios $R^{\text{incl}}$ and $R^{\mathrm{MN}}$ in a wide range of rapidity separation, up to $\abs{\Delta y} = 9.2$,
in proton-proton collisions at $\sqrt{s}=7\TeV$ was presented.
A moderate rise of the ratio of the inclusive to exclusive dijet production cross sections as a function of
$\abs{\Delta y}$ is observed. The predictions of the \PYTHIA{}6 and \PYTHIA{}8
generators agree with the measurements. The predictions of the \HERWIGpp generator are larger than the measurement especially at large $\abs{\Delta y}$. The BFKL-motivated
generators \textsc{cascade} and \textsc{hej+ariadne} predict for these ratios a
significantly stronger rise than observed.
The moderate rise of the measured dijet ratios indicates that the BFKL effects are not dominant for jets with $\pt > 35\GeV$ at the present collision energy of $7\TeV$.

We congratulate our colleagues in the CERN accelerator departments for the excellent performance of the LHC machine. We thank the technical and administrative staff at CERN and other CMS institutes, and acknowledge support from: FMSR (Austria); FNRS and FWO (Belgium); CNPq, CAPES, FAPERJ, and FAPESP (Brazil); MES (Bulgaria); CERN; CAS, MoST, and NSFC (China); COLCIENCIAS (Colombia); MSES (Croatia); RPF (Cyprus); MoER, SF0690030s09 and ERDF (Estonia); Academy of Finland, MEC, and HIP (Finland); CEA and CNRS/IN2P3 (France); BMBF, DFG, and HGF (Germany); GSRT (Greece); OTKA and NKTH (Hungary); DAE and DST (India); IPM (Iran); SFI (Ireland); INFN (Italy); NRF and WCU (Korea); LAS (Lithuania); CINVESTAV, CONACYT, SEP, and UASLP-FAI (Mexico); MSI (New Zealand); PAEC (Pakistan); MSHE and NSC (Poland); FCT (Portugal); JINR (Armenia, Belarus, Georgia, Ukraine, Uzbekistan); MON, RosAtom, RAS and RFBR (Russia); MSTD (Serbia); MICINN and CPAN (Spain); Swiss Funding Agencies (Switzerland); NSC (Taipei); TUBITAK and TAEK (Turkey); STFC (United Kingdom); DOE and NSF (USA).

\bibliography{auto_generated}   

\cleardoublepage \appendix\section{The CMS Collaboration \label{app:collab}}\begin{sloppypar}\hyphenpenalty=5000\widowpenalty=500\clubpenalty=5000\textbf{Yerevan Physics Institute,  Yerevan,  Armenia}\\*[0pt]
S.~Chatrchyan, V.~Khachatryan, A.M.~Sirunyan, A.~Tumasyan
\vskip\cmsinstskip
\textbf{Institut f\"{u}r Hochenergiephysik der OeAW,  Wien,  Austria}\\*[0pt]
W.~Adam, T.~Bergauer, M.~Dragicevic, J.~Er\"{o}, C.~Fabjan, M.~Friedl, R.~Fr\"{u}hwirth, V.M.~Ghete, J.~Hammer\cmsAuthorMark{1}, M.~Hoch, N.~H\"{o}rmann, J.~Hrubec, M.~Jeitler, W.~Kiesenhofer, M.~Krammer, D.~Liko, I.~Mikulec, M.~Pernicka$^{\textrm{\dag}}$, B.~Rahbaran, C.~Rohringer, H.~Rohringer, R.~Sch\"{o}fbeck, J.~Strauss, A.~Taurok, F.~Teischinger, P.~Wagner, W.~Waltenberger, G.~Walzel, E.~Widl, C.-E.~Wulz
% ...
\vskip\cmsinstskip
\dag:~Deceased\\
1:~~Also at CERN, European Organization for Nuclear Research, Geneva, Switzerland\\
% ...

\end{sloppypar}
\end{document}